\input{aipcheck}

\documentclass[
    ,final            
  ]
  {aipproc}

\layoutstyle{6x9}
\usepackage{wrapfig}

\begin{document}

\title{Constraining pulsar gap models with the light-curve and flux properties of the gamma-ray 
pulsar population}


\author{Marco Pierbattista}{
  address={Laboratoire AIM, CEA-IRFU/CNRS/ Paris Diderot University, Service dAstrophysique, 
  CEA Saclay, 91191 Gif sur Yvette, France \& Paris Diderot University}
}

\author{Isabelle Grenier}{
  address={Laboratoire AIM, CEA-IRFU/CNRS/ Paris Diderot University, Service dAstrophysique, 
  CEA Saclay, 91191 Gif sur Yvette, France \& Paris Diderot University}
}

\author{Alice Harding}{
  address={NASA Goddard Space Flight Center, Greenbelt, MD 20771, USA}
}

\author{Peter L. Gonthier}{
  address={Hope College, Department of Physics, MI 49423, USA}
}

\begin{abstract}
 We compare population synthesis results for inner and outer magnetosphere emission models 
 with the various characteristics measured in the first  LAT pulsar catalogue for both the radio-loud 
 and radio-weak or radio-quiet $\gamma$-ray pulsars. 
 We show that all models fail to reproduce the observations: for each model there is a lack of luminous 
 and energetic objects that suggest a non dipolar magnetic field structure or spin-down evolution.
 The large dispersion that we 
 find in the simulated gamma-ray luminosity versus spin-down power relation 
 does not allow to use the present trend seen in the Fermi data to distinguish 
 among models. For each model and each Fermi detected pulsar, we have generated light curves as 
 a function of obliquity and inclination angles. The theoretical curves were fitted to the observed one, 
 using a maximum-likelihood approach, to derive the best-fit orientations and to compare how well 
 each model can reproduce the data. Including the radio light-curve gives an 
 additional key constraint to restrict the orientation space. 
 \end{abstract}

\maketitle


\section{ Radio pulsar population synthesis and emission models}
We synthesised a population of $10^7$ neutron stars (NSs). The stars were born on the galactic plane with a surface 
density distribution obtained from the location of the HII clouds in the Milky Way and spread in the Galaxy by assuming 
a maxwellian supernova kick velocity distribution with mean of 400 kms$^{-1}$. Each star has been spun down as a rotating
dipolar magnetic field and by assuming a constant exponential rate magnetic field decay on a timescale of 2.8 Myr. 
For the details and references on the NS sample evolution see the Gonthier et al. proceeding of this same conference. 

\subsection{Gamma-ray and radio emission models}
For each NS of the simulated sample we have evaluated $\gamma$-ray luminosity and emission pattern 
according to 4 different emission models: Polar Cap, PC (low magnetosphere, two pole caustic) \cite{PCmh03}, 
Slot Gap, SG (intermediate-high magnetosphere, two pole caustic) \cite{SGmh04}, Outer Gap (OG) \cite{OGcrz00}
and One Pole Caustic (OPC) \cite{OPCrw10} (both high magnetosphere, one pole caustic).
The radio emission from each NS of the sample has been simulated by assuming an emission cone composed
by a core and cone components \cite{ hgg04}.

\subsection{Gamma \& radio visibility}
For each assumed emission model, the visible component of the whole pulsar population has been obtained by 
applying a one year LAT pulsar visibility map. Each $\gamma$-ray visible subsample has then been scaled to the number
of radio pulsars in 10 radio surveys of the ATNF catalogue. This gives the number of simulated visible pulsars that would 
have been observed by the LAT telescope on the FERMI satellite after one year of observation, by assuming each emission 
model.

\section{Fitting the pulsar emission with the implemented models}
By using the emission pattern of the implemented emission model, we performed an estimate of the magnetic obliquity $\alpha$
(angle between magnetic and spin axis) and line of sight $\zeta$ (angle between spin axis and observer direction) for a sample of 
22 radio loud (RL) LAT pulsars. The estimate has been implemented by fitting jointly the $\gamma$-ray and radio light curves. 
A second $(\alpha ; \zeta)$ estimate has been performed by fitting just the $\gamma$-ray LAT pulsar light curves with the high-energy
model emission patterns. The comparison between the $(\alpha ; \zeta)$ solution space of the joint radio+gamma and gamma fit
will be discussed in the results section. In both the fit methods and for each LAT pulsar, we have found a fit solution for each one
of the assumed emission models. For each fitted pulsar, the model fit solutions (PC, SG, OG or OPC) characterised by the highest 
significance (highest likelihood value), is called \emph{best fit solution}.

\section{Results and discussion}
\begin{figure}[htb!]
  \includegraphics[width=0.44\textwidth]{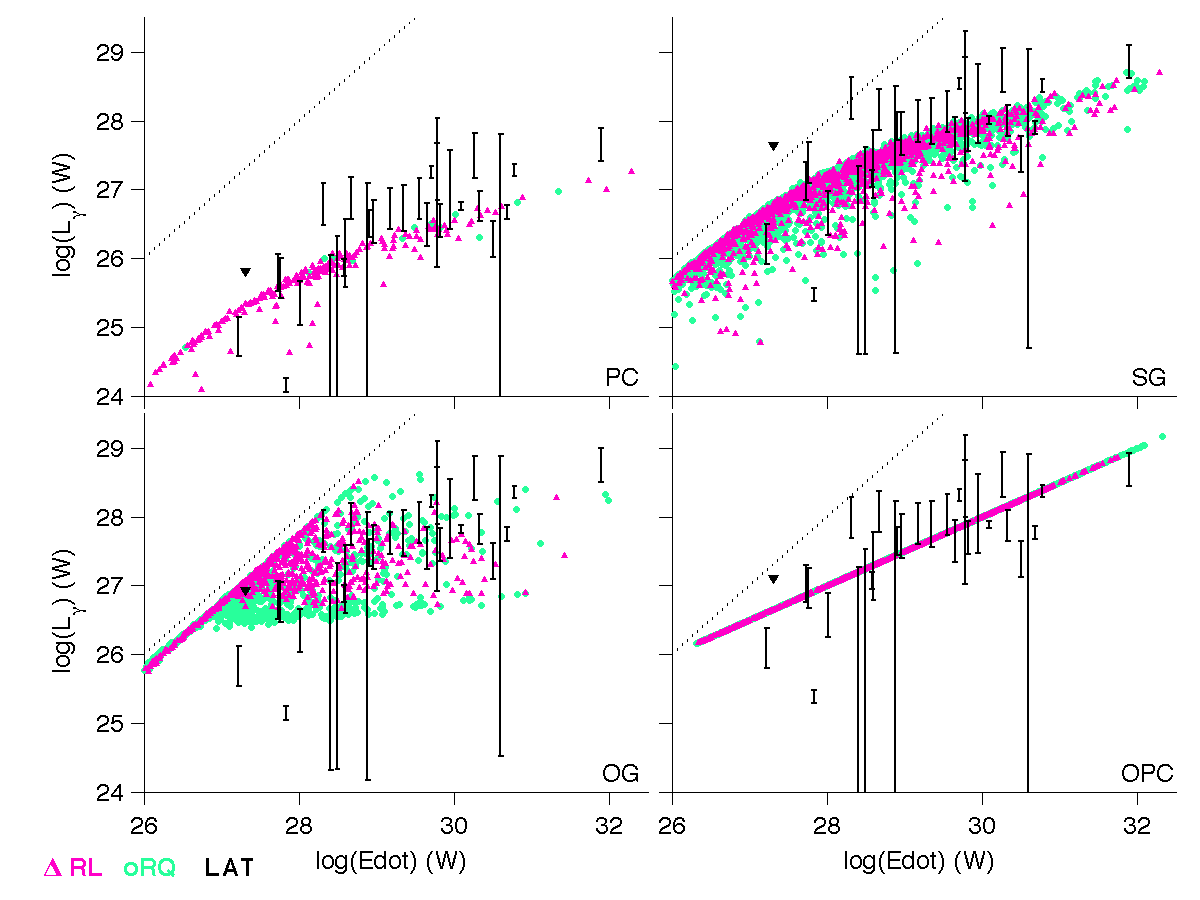}
  \includegraphics[width=0.44\textwidth]{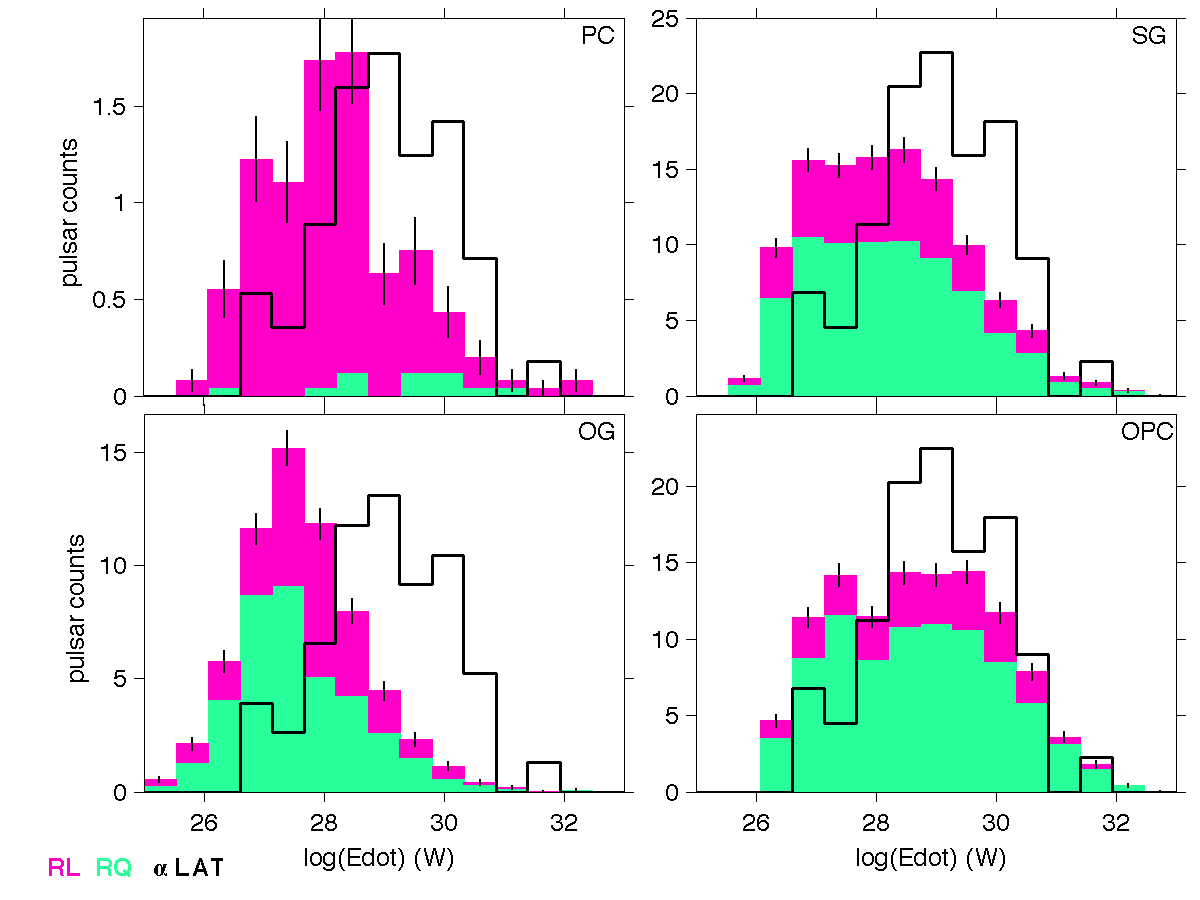}
  \caption{\footnotesize \emph{left}: for each model is shown the comparison between  $L_{\gamma}=f(\dot{E})$ simulated (colored)  
                   and observation (black). \emph{right:} $\dot{E}$ distribution comparisons between data (black) and simulations 
                   In both the plots, pink and green respectively refer to RL \& RQ simulated pulsars.}
  \label{f1}
\end{figure}

\paragraph{Comparison LAT observations model predictions}
In figure \ref{f1} are shown the comparisons between the LAT observations and simulations, for luminosity vs. spin-down luminosity (left) 
and spin-down luminosity distribution (right). The PC, SG, and OPC model predictions well describe the expected trend 
$L_{\gamma}\propto \dot{E}^{0.5}$ while the OG model shows a too high dispersion. All the models show a lack of 
luminous objects and so fail in describing the observed population. The gamma luminosity comparison cannot be used to 
discriminate which model best describe the observations.
Figures \ref{f1} right shows that there is a lack of energetic objects in the model predictions. No one of the proposed models is able to explain
the high number of high $\dot{E}$ pulsars observed by the LAT. The lack is obviously non model dependent and is not due to the choice
of intrinsic parameters of the stars, like radius, mass or moment of inertia. Several configuration of these intrinsic parameters were tried and all of
them led to the same lack of energetic objects. The current mass, radius, and moment of inertia configuration has been chosen to boost the power
and minimise the observed lack. We identify two most probable explanations to the lack of energetic objects. A first one is based on a different magnetic
field structure (non dipolar) and so spin-down luminosity evolution for very energetic objects. A second one hypothesises a magnetic obliquity $\alpha$ evolution during the 
first stages of a pulsar life. 
The different shape between simulated and observed $\dot{E}$ distributions (fig. \ref{f1} right) could support the non dipolar spin-down luminosity evolution
scenario for the most energetic pulsars. The fact that the best agreement between data and simulations is obtained for the OPC model and since it differs from the
OG just for the dependency of the luminosity from $\alpha$ (OG: $L_{\gamma}=f(\dot{E},\alpha)$; OPC: $L_{\gamma}=f(\dot{E})$), could be one element in favor
of the the magnetic obliquity evolution scenario and so, a possible $\alpha$ evolution for energetic objects should be taken into account.

\begin{figure}[htb!]
  \includegraphics[width=0.44\textwidth]{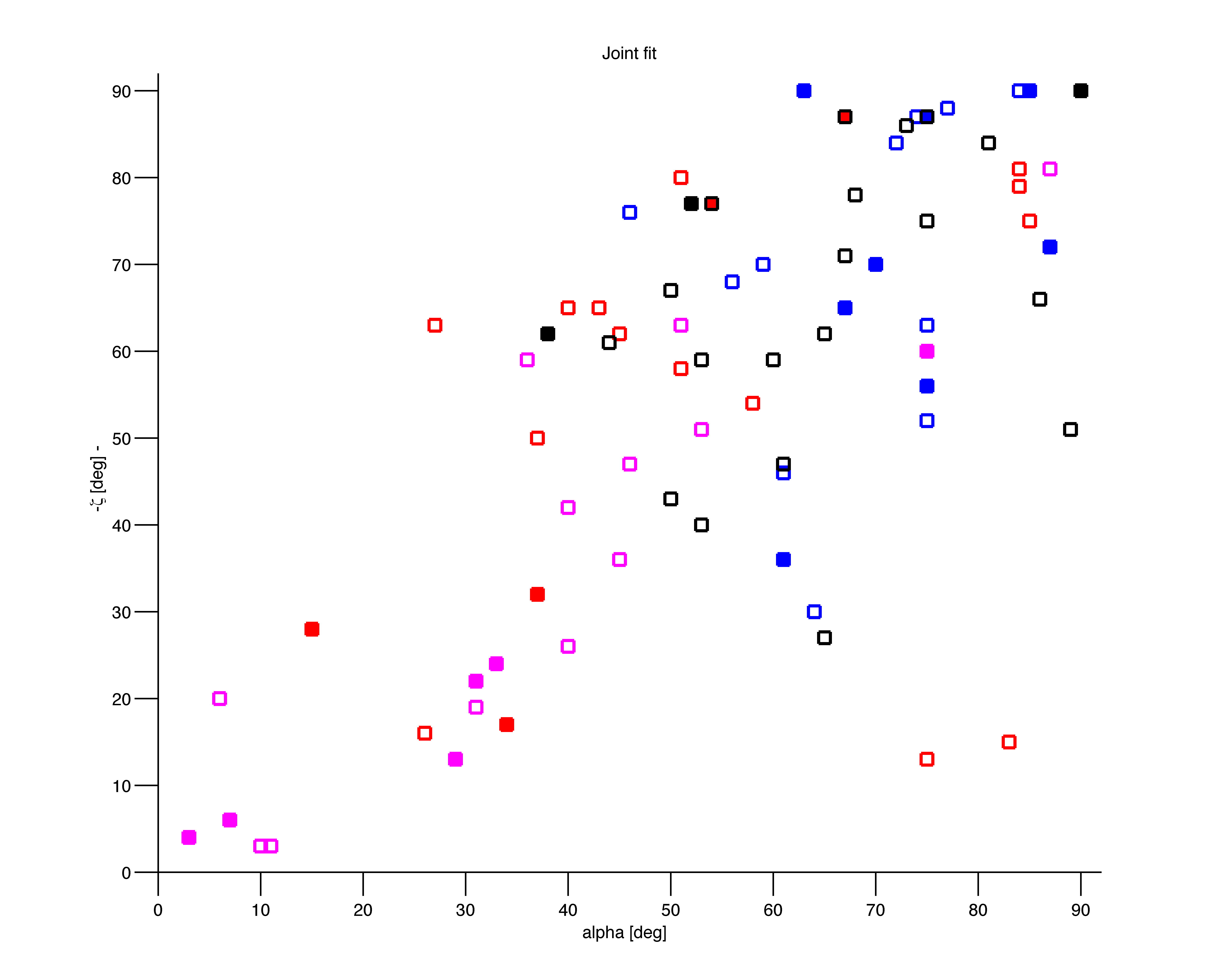}
   \includegraphics[width=0.44\textwidth]{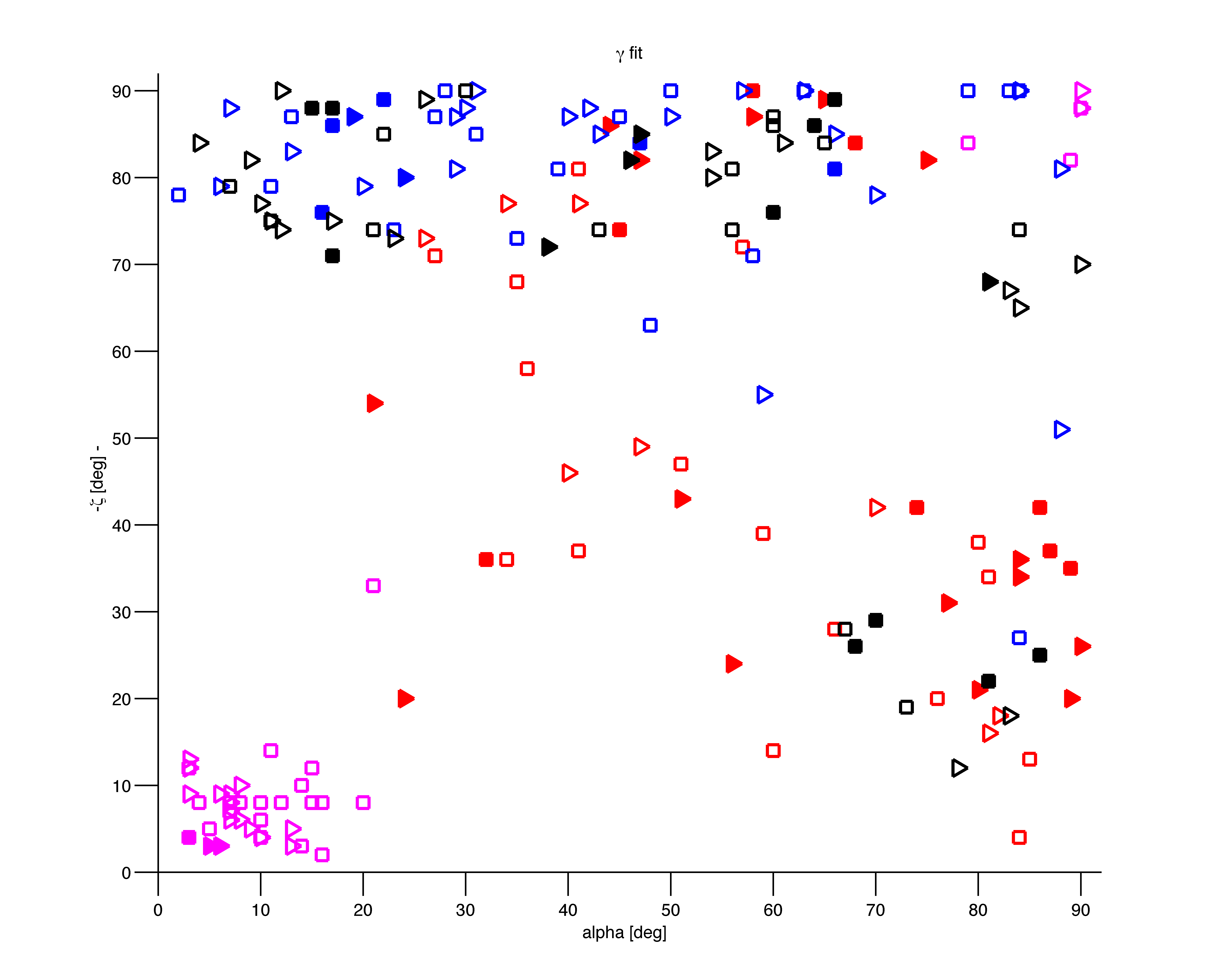}
  \caption{\footnotesize $(\alpha ;\zeta)$ solutions plane for the joint radio plus gamma fit results (left) and just gamma results (right).  
                  In the right plot, triangles and square respectively refer to radio quite (RQ) and radio loud (RL) pulsars. In both the plots, pink, red, blue
  	         and black refer to PC, SG, OG, and OPC model solutions, and empty and filled markers respectively refers to all the solutions and to 
	         the best fit ones.}
  \label{f2}
\end{figure}
\begin{figure}[htb!]
   \includegraphics[width=0.45\textwidth]{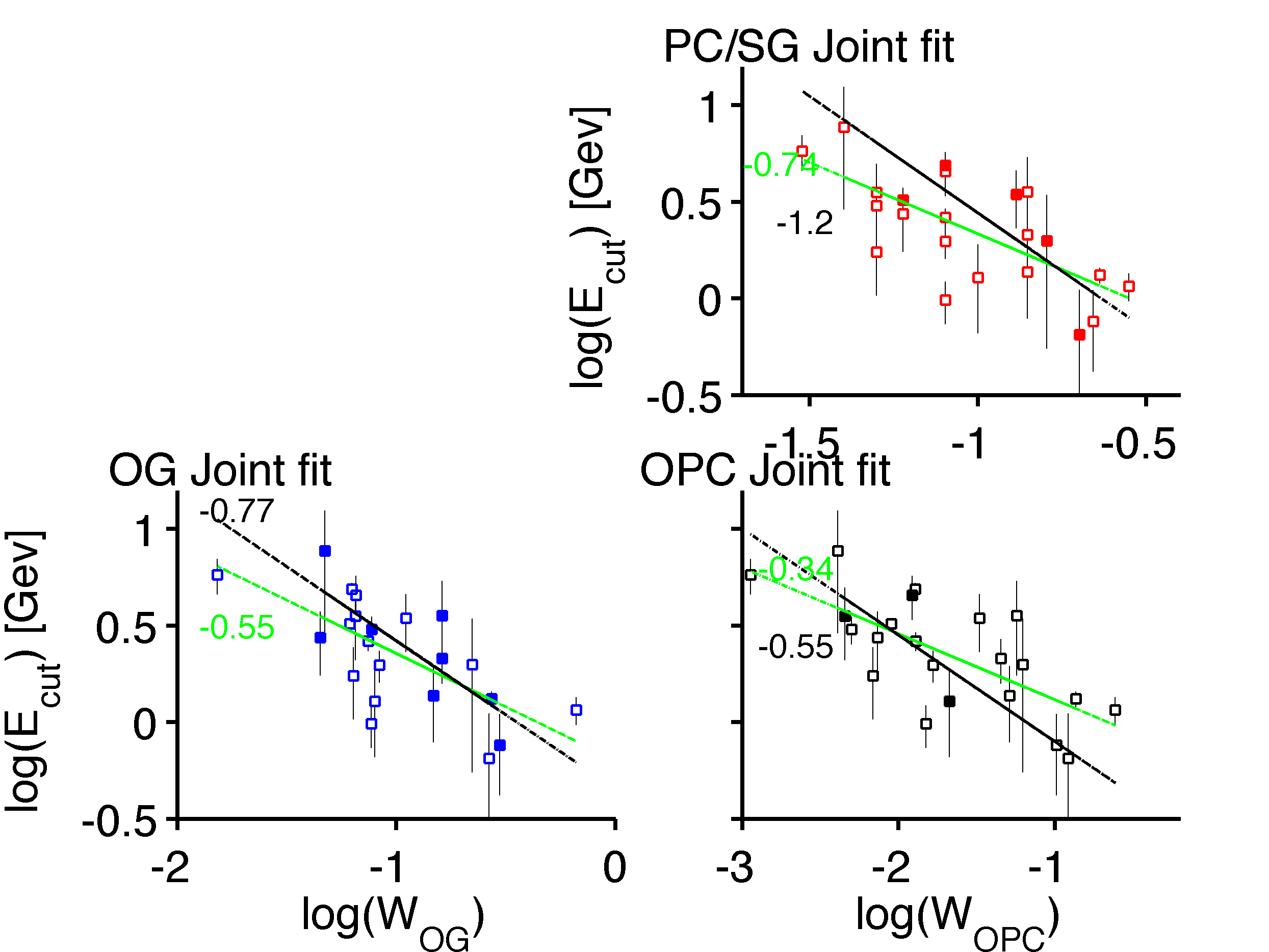}
  \caption{\footnotesize High energy cutoff as a function
                 of the accelerator gap width. Green and black lines respectively refer to the best fit power law for all the solutions and for the highest
                 significance ones. In both the plot, empty and filled markers respectively refers to all the solutions and to the highest significance ones.}
  \label{f3}
\end{figure}

\paragraph{Fit of the LAT pulsar profiles}
In figure \ref{f2} and \ref{f3} are shown two of the main and suggestive results of the joint fit estimate of the LAT pulsars orientation. 
Figure \ref{f2} shows the comparison between the $(\alpha; \zeta)$ solutions plane for the joint radio plus gamma and just gamma fit.
The gamma fit solution plane (figure \ref{f2} right) contains RL solutions in a region of the plane where is not possible to have radio emission.
As it is shown in figure \ref{f2} left panel, to be able to explain both the radio and gamma emission, the RL pulsar, can have $(\alpha; \zeta)$
solution just along the radio diagonal of the $\alpha -\zeta$ plane ($\alpha; \zeta=0; 0 \rightarrow \alpha; \zeta=90; 90$). This is justified by the
fact that to be able to observe simultaneously radio and gamma emission, the angular distance between the observer line of sight $\zeta$ and the
magnetic axis can not be bigger of the angular radius of the radio emission beam that, for young pulsars, assume the values of $\sim 20-30$ degrees.
  
In figure \ref{f3} is shown the relation between the high energy cutoff and the gap width we found, for each model, in the fit results and not in the 
population study. This relation is particularly important because connect an intrinsic pulsar
parameter like the with of the accelerator gap and the observed high energy cutoff. From the SG models, we obtained that $E_{cut}\propto w_{gap}^{-0.5}$.

\paragraph{Conclusions}
The comparison between luminosity evolution and $\dot{E}$ distribution of simulated and observed samples show a lack of energetic objects
in the model predictions. As a possible cause, different spin-down luminosity evolution/magnetic field structures and/or $\alpha$ evolution have to be more investigated.
The comparison between the $(\alpha ;\zeta)$ plane solutions of the joint fit and gamma fit method suggest that a multi-wavelength pulsar light curve fit is the only
way to obtain reliable  $(\alpha ;\zeta)$ estimates.
The energy cutoff-gap width relation found offers the possibility for a future improvement of the population synthesis.

\end{document}